\documentclass[conference]{IEEEtran}
\usepackage{url}
\usepackage{tikz}
\usepackage{amsmath}
\usepackage[T1]{fontenc}
\usepackage[utf8]{inputenc}
\usepackage{graphicx}
\usepackage{epsfig}
\usepackage{amsmath}
\usepackage{bbold}
\usepackage{color}
\usepackage[table]{xcolor}
\usepackage[ruled,vlined]{algorithm2e}
\usepackage{subcaption}
\usepackage{wrapfig}
\usepackage{booktabs}
\usepackage{pifont}
\usepackage{xspace}
\usepackage{cite}
\usepackage{breakurl}
\usepackage{hyperref}
\usepackage{graphicx}
\usepackage{booktabs}
\usepackage{adjustbox}
\usepackage[normalem]{ulem}
\hypersetup{
  colorlinks,
  linkcolor={green!80!black},
  citecolor={red!70!black},
  urlcolor={blue!70!black}
}
\usepackage{amssymb}

\graphicspath{{./figures/}}

\newtheorem{proposition}{\textbf{Proposition}}

\newcommand{\sysname}{DFL-C\xspace}

\newcommand{\tsname}{D2TS\xspace}

\title{Model-Consistent Byzantine-Resilient Decentralized Federated Learning for Collaborative Missions}

\author{
\IEEEauthorblockN{
Yue Li\IEEEauthorrefmark{1}, 
Sudip Bhujel\IEEEauthorrefmark{1},
Cameron Lira\IEEEauthorrefmark{1},
Ning Wang\IEEEauthorrefmark{2},
Yang Xiao\IEEEauthorrefmark{1}
}
\IEEEauthorblockA{\IEEEauthorrefmark{1}University of Kentucky, Lexington, KY, USA}
\IEEEauthorblockA{\IEEEauthorrefmark{2}University of South Florida, Tampa, FL, USA}
}

\begin{document}
\maketitle

\begingroup
\renewcommand\thefootnote{}
\footnotetext{
This is the authors' final version of the paper, which has been accepted for publication at the IEEE Conference on Communications and Network Security (CNS), 2026.
}
\addtocounter{footnote}{-1}
\endgroup

\begin{abstract}

Decentralized federated learning (DFL) is a promising paradigm for autonomous nodes to collaboratively train AI models without relying on a central server. However, existing DFL solutions do not guarantee global model consistency, a critical requirement for collaborative mission-critical scenarios where model divergence undermines decision uniformity and safety. This lack of consistency also amplifies vulnerability to Byzantine adversaries, who exploit the decentralized network topology and weak synchrony to perform equivocation and model poisoning attacks against individual victims.

This paper introduces DFL-C, a novel Byzantine-resilient DFL architecture that enables decentralized nodes to perform collaborative training with global model consistency. At its core, DFL-C integrates an asynchronous common subset (ACS) consensus protocol into the DFL workflow to ensure all nodes aggregate a uniform set of model updates to establish global model consistency, despite individual Byzantine equivocation. DFL-C further implements a dual-domain trust scoring mechanism to provide resilience against data-domain Byzantine manipulations including model poisoning attacks. This mechanism complements the consensus protocol, significantly reducing the latter's runtime. Our experimental results demonstrate that DFL-C maintains model accuracy while achieving global model consistency under Byzantine behaviors with moderate consensus overhead. Notably, when compared with the state-of-the-art DFL solution BALANCE (Fang et al.) that does not provide model consistency, DFL-C achieves better model accuracy against untargeted model poisoning attacks and comparable resilience against backdoor attacks, with the advantage widened under non-IID scenarios.



\end{abstract}



\section{Introduction}
\label{sec:intro}

Decentralized federated learning (DFL) \cite{vanhaesebrouck2017decentralized,lalitha2018fully,beltran2023decentralized} is a robust paradigm for collaborative model training across autonomous participants. It provides a unique advantage for AI and machine learning tasks in a decentralized, mission-critical setting where maintaining a stable, trusted leader is infeasible, with promising use cases in joint sensing and swarm robotics \cite{wang2020learning,nguyen2022deep,zhou2023decentralized}.
Unlike classic FL \cite{mcmahan2017communication} which relies on a central server to orchestrate training, DFL enables participants to self-organize the iterative learning process, eliminating the need for a central orchestrator \cite{qu2022decentralized}. 
In a typical setting, DFL participants form a peer-to-peer network to iteratively exchange locally trained model updates perform independent aggregation, until each local model achieves convergence.
While offering robust decentralization and resilience to centralized server failures, established peer-to-peer DFL solutions \cite{vanhaesebrouck2017decentralized,lalitha2018fully,beltran2023decentralized} often forfeit \emph{global model consistency}, a property that is guaranteed in classic FL. They mandate all participants to update local model by aggregating over updates only from neighbors; local models across the network will gradually diverge over successive iterations. The divergence happens even faster between topologically distant nodes 
\cite{shi2023improving}. For a collaborative AI-driven mission, model divergence can undermine network-wide decision integrity and safety \cite{qu2022decentralized}. For instance, in a joint sensing mission, inconsistent models across the participants may lead to conflicting classifications of a target, resulting in mission failures. Achieving model consistency in a decentralized setting would demand a form of consensus that unifies the trained model across the network. 



Besides the model divergence problem, localized aggregation in peer-to-peer DFL is inherently more prone to adversarial (\emph{Byzantine}) influence. 
First, Byzantine participants may equivocate, by sending different model updates 
to different peers, to create performance gaps across the network. Equivocation is difficult to detect at the local level if the DFL participants only hear from their neighbors. 
Second, prior studies show that Byzantine participants can leverage model poisoning attacks (MPAs) \cite{tolpegin2020data,bhagoji2019analyzing,gu2019badnets,fang2020local,fowl2021robbing,nguyen2023iba} to manipulate model aggregation at each neighbor.
Compared to classic FL where the server conducts central aggregation, localized aggregation in DFL is more vulnerable to MPAs
since each participant aggregates over its neighbors’ updates, not the global set of updates. An adversary may easily control the majority of a victim's neighbors, even if the compromised peers constitute only a small portion of the overall network.

Last but not the least, existing peer-to-peer DFL solutions typically assume synchronous model updates, where each participant receives inputs from all (or a certain portion of) neighbors in each round. This is often infeasible in decentralized mission-critical networks due to systematic asynchrony that arises from two factors: sporadic network conditions leading to volatile communication delays and heterogeneous training speeds among participants leading to asynchronous model updating. 
While recent works have explored asynchronous mechanisms in DFL \cite{bornstein2022swift,liu2024aedfl,jeong2025draco}, they either do not tolerate full asynchrony (no predictable bound for model updating delay) or do not enforce model consistency.
\textbf{Contributions.}
We introduce \textbf{\sysname}, a novel DFL architecture designed for the collaborative training of a shared global model among autonomous participants (hereafter referred to as \emph{nodes}). \sysname is especially useful for model training in a mission-critical network where nodes do not have an a priori trust hierarchy, operate over intermittent connectivity, and require consistent model-guided decisions for operational safety. 
The architecture is resilient to Byzantine behaviors, including equivocation and major MPA types, and attains inference accuracy comparable to centralized FL benchmarks. Furthermore, \sysname tolerates full asynchrony in both communication and model updates.
To achieve these objectives, \sysname incorporates three design components that unfold incrementally.

\begin{figure}
    \centering
    \begin{minipage}{\columnwidth}
        \begin{subfigure}{0.3\textwidth}
            \includegraphics[width=\linewidth]{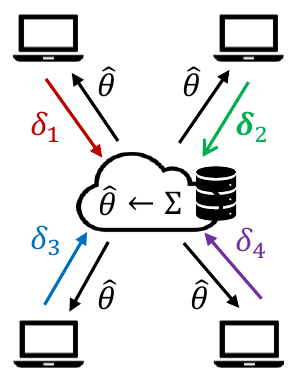}
            \caption{Classic FL}
            \label{fig:model-classic-fl}
        \end{subfigure}
        \hfill
        \begin{subfigure}{0.3\textwidth}
            \includegraphics[width=\linewidth]{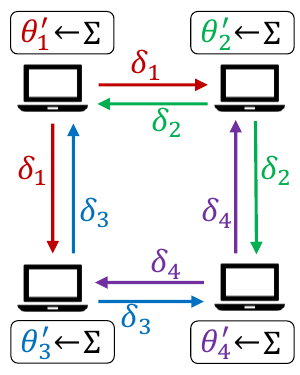}
            \caption{DFL}
            \label{fig:model-dfl}
        \end{subfigure}
        \hfill
        \begin{subfigure}{0.3\textwidth}
            \includegraphics[width=\linewidth]{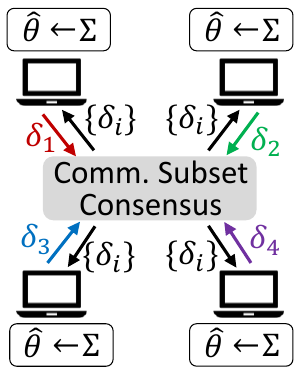}
            \caption{DFL-C}
            \label{fig:model-dflc}
        \end{subfigure}
    \end{minipage}
    \caption{Comparing classic FL, existing (peer-to-peer) DFL  and \sysname high-level workflow in one training round.}
    \label{fig:comparing-three-FL-models}
\end{figure}

\textbf{(1) Integrating common subset consensus into DFL:} \sysname builds on the core intuition that the peer-to-peer DFL paradigm can integrate common subset consensus, a special form of Byzantine fault-tolerant (BFT) consensus, to enforce global model consistency. 
As shown in Fig. \ref{fig:model-dflc}, every \sysname round begins with nodes training on the current global model $\hat{\theta}$ with local data, then proposing their resulting updates to the network. Through the common subset consensus, all honest nodes agree on a common subset of updates, denoted $\{\delta_i\}$. 
Under the assumption that $N\geq 3F+1$ ($N$ is the total number of nodes and $F$ the Byzantine nodes), the common subset is guaranteed to contain at least $N-F$ model updates with the majority of which originated from honest nodes---allowing \sysname to attain model accuracy comparable to centralized FL.
This one-third threshold is critical for mitigating equivocation: with at most $F$ Byzantine nodes, any two quorums of size $N-F$ intersect in at least $F+1$ nodes, so their overlap contains at least one honest node, allowing common subset consensus to exclude conflicting updates from any equivocating node, while still making progress.
Finally, all honest nodes apply a deterministic aggregation algorithm over the common subset of updates to derive an identical new model $\hat{\theta}$. 



\textbf{(2) Dual-domain trust Scoring for adaptive Byzantine resilience:} To mitigate adaptive Byzantine behaviors that the adversary perform equivocation and MPA attacks over time, \sysname further adopts a dual-domain trust scoring scheme, dubbed \textbf{\tsname}, to be executed by honest nodes. It generates a uniform aggregation weight vector over the updates in the common subset which is then used for weighted local aggregation. 
\tsname evaluates node reliability across two distinct operational domains: the consensus domain, which penalizes equivocation and non-responsiveness, and the data domain, which evaluates model update quality. Specifically, the update quality factors in spatial consistency (closeness to other updates in parameter space) and temporal consistency (stability of updates across successive rounds), distinguishing honest updates from low-quality or MPA updates. Therefore, \tsname enables \sysname to react dynamically to Byzantine behaviors, mitigating their impact on the accuracy of the final model. 


\textbf{(3) Tolerating asynchrony while improving consensus efficiency:} To tolerate systematic asynchrony typically seen in a decentralized network, \sysname adapts the asynchronous common subset (ACS) framework \cite{miller2016honey,guo2020dumbo,duan2023fin} for the aforementioned consensus function.
As a concrete ACS realization, we design \emph{trust-informed ACS} (\textbf{T-ACS}) that utilizes the trust scores generated by \tsname in the preceding round to inform the prioritization of higher-scored model updates in the current consensus session. It accelerates the common subset consensus by proactively filtering out low-priority updates without compromising accuracy, reducing the substantial latency typically associated with a ACS scheme. Another important function of T-ACS is to capture equivocation signals during the consensus and feeds them to \tsname for trust scoring.

Lastly, for practical efficiency when training time is predictable, \sysname integrates an adaptive waiting policy to opportunistically mitigate the unnecessary consensus runtime caused by heterogeneous training times. It allows faster nodes (who finish local training quickly) delay the start of their T-ACS session to align with the network's pace. This policy minimizes the time that fast nodes spend idle in consensus, preserving computational resources for other tasks or enabling power-saving modes.

\textbf{Evaluation highlights.} We evaluated \sysname in networks ranging from 4 to 13 nodes tasked with training the LeNet \cite{lecun1998gradient} and ResNet \cite{he2016deep} classification models. Results show that \sysname achieves global model consistency and equivocation resistance while incurring moderate consensus runtime, minimal when compared to local training time, in a decentralized node cluster. We compared \sysname with BALANCE \cite {fang2024byzantine}, the state-of-the-art Byzantine-resilient DFL solution designed to counter MPAs but not providing model consistency.
Under Byzantine equivocation and label flipping \cite{tolpegin2020data}, a popular untargeted MPA, \sysname's model at convergence attains higher accuracy than BALANCE, in addition to achieving model consistency.
Notably, the model accuracy gap between \sysname and BALANCE widens when the local data is more non-independent and identically distributed (non-IID).
\section{Background and Related Work}
\label{sec:background}







\subsection{Decentralized Federated Learning}

Established DFL formulations \cite{vanhaesebrouck2017decentralized,lalitha2018fully,beltran2023decentralized} adopt the peer-to-peer model without relying on a central server. This structure enables direct communication between adjacent nodes that perform local model aggregation with inputs from neighbors. 
A DFL round involves a two-step procedure as shown in Fig. \ref{fig:model-dfl}. First, each node $i$ performs local training on its local model $\theta_i$ over its own data to generate a local model update $\delta_i$ before sending it to peers. Second, each node $i$ performs local aggregation on all received model updates (including its own) to derive the new model $\theta_i'$. This paradigm, however, allows the new models to differ among the nodes. While model heterogeneity is beneficial in static cross-silo FL scenarios, it imposes utility or safety risks in mission-critical scenarios, where model consistency is important \cite{qu2022decentralized}.
Alternatively, semi-decentralized FL achieves model consistency by relying on a rotating aggregator \cite{lin2021semi,yemini2022semi}.
or a blockchain platform \cite{li2020blockchain} for uniform model aggregation. 
However, they are not fully decentralized due to the rotating leader or the blockchain platform being a new aggregation bottleneck.

To handle asynchrony of model updates, recent works have explored asynchronous DFL formulations  \cite{liu2024aedfl,bornstein2022swift,jeong2025draco}. AEDFL \cite{liu2024aedfl} needs to assume bounded model updating delays (i.e., weak synchrony only). SWIFT \cite{bornstein2022swift} and DRACO \cite{jeong2025draco} tolerate full asynchrony but do not enforce model consistency.

\subsection{Byzantine Attacks and Defense}

In classical FL, malicious (Byzantine) clients can craft model updates to manipulate the central aggregation through model poisoning attacks (MPA) \cite{tolpegin2020data,bhagoji2019analyzing,gu2019badnets,fang2020local,fowl2021robbing,nguyen2023iba}. MPAs fall into two classes: untargeted attacks, which aim to degrade the overall model accuracy, and targeted (backdoor) attacks, which aim to corrupt the model's performance on specific sub-tasks without affecting its performance on the main task. Traditional Byzantine-resilient aggregation rules (BRARs) 
\cite{blanchard2017machine,yin2018byzantine} are shown to have limited efficacy in handling MPAs, 
especially the backdoor variants.
To enhance MPA robustness, recent methods compute a trust score for each client by measuring its model update's deviation from the majority in a specific parameter space \cite{cao2020fltrust,zhang2022fldetector,wang2022flare,fereidooni2023freqfed}, or temporal stability \cite{krauss2023mesas}. 

However, byzantine-resilient aggregation techniques designed for classic FL have limited efficacy in DFL, mainly because in DFL, nodes aggregate updates from a limited group of neighbors rather than the global population \cite{beltran2023decentralized}. Any global-level Byzantine threshold assumption (e.g., one-third) may be violated locally, especially for a targeted victim whose neighbors be can predominantly malicious. Moreover, DFL is uniquely vulnerable to equivocation attacks, where Byzantine nodes send conflicting updates to different peers to create performance gaps across the network \cite{shi2023improving}.

A few Byzantine resilience schemes have been proposed for DFL \cite{el2021collaborative,guo2021byzantine,he2022byzantine,fang2024byzantine} with the intuition that each participant should be selective on its neighbor inputs to contain the influence of Byzantine peers. 
Specifically, \textsc{ClippedCossip}~\cite{he2022byzantine} is the first DFL algorithm to achieve global model consistency; however, it does not handle MPAs and its consistency mechanism does not handle Byzantine equivocation. 
BALANCE~\cite{fang2024byzantine} is the state-of-the-art Byzantine-robust DFL scheme by using a trust-scoring mechanism for localized defense against known MPAs, providing a proven convergence rate guarantee.
However, it does not maintain a consistent model across decentralized nodes or consider an adversary who equivocates or targets a victim's neighborhood.

\begin{figure*}
    \centering
    \includegraphics[width=.9\textwidth]{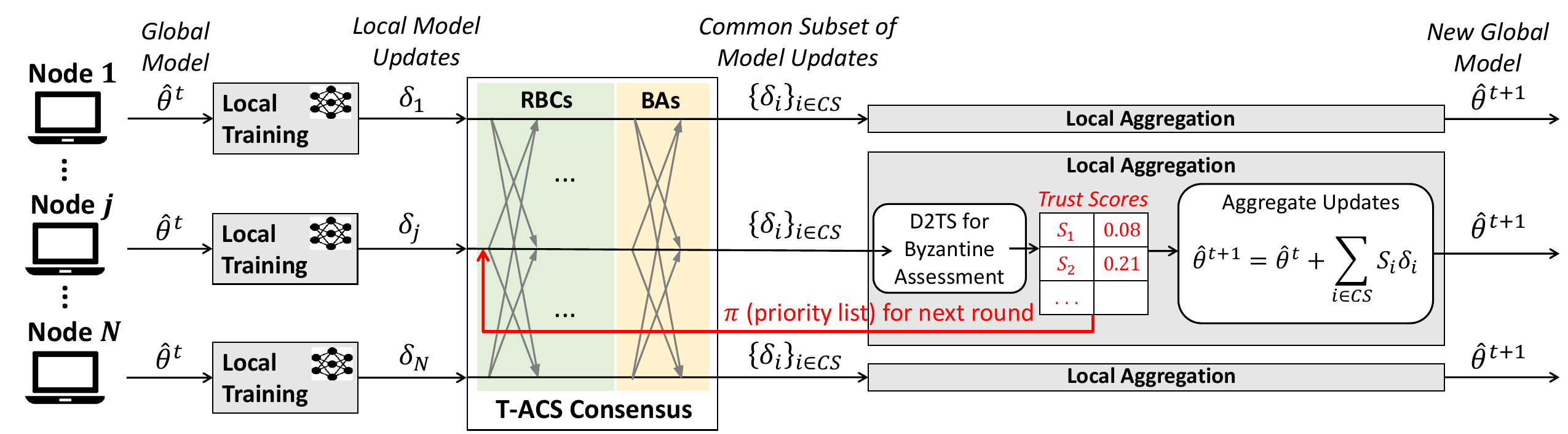}
    \caption{\sysname workflow in one training round.}
    \label{fig:sys-arch}
\end{figure*}

\section{System Model}


\subsection{Network Model}

We consider a network of $N$ independent nodes, where any pair of nodes can communicate via an \textit{authenticated} but \textit{asynchronous} channel. The assumption of message authentication is established in the mobile Internet and increasingly in IoT systems, where standard transport-layer or application-layer security mechanisms are available. The asynchrony assumption models the worst-case scenario where messages may suffer from arbitrary delays; however, they are guaranteed to arrive eventually. Together, the two assumptions provide a conservative yet practical model for a collaborative mission-critical network where nodes do not have an a priori trust hierarchy and operate over intermittent connectivity while being deployed to a common mission.


\textbf{Task Model.} 
We consider DFL tasks where the $N$ nodes jointly train a machine learning model. Each node accumulates training data samples locally. For supervised learning tasks, we assume the data samples are labeled by local resources. While our DFL framework can be extended to unsupervised cases, we focus on supervised learning in this work. Nodes carry out the FL task on a \emph{round} basis---for each round, they start from the same initial model $\theta$ (result from the previous round) and their respective local datasets.
The round's task is to retrain $\theta$ into a new common model. 



\subsection{\sysname Objectives}

\textbf{Global model consistency.} All honest nodes should obtain the same model at a round's end.

\textbf{Accuracy.} The converged global model should maintain a level of inference accuracy on par with centralized FL.

\textbf{Byzantine resilience.} Even in the presence of Byzantine nodes performing equivocation or model manipulation attacks (see Threat Model), the above objectives are still guaranteed.

\textbf{Tolerating asynchrony.} The asynchrony in communication and model updating should not degrade the above objectives.

While achieving these objectives, a practical solution should minimize its communication overhead if possible.

\subsection{Threat Model}
\label{subsec:threat-model}

We assume that up to $F$ out of $N$ nodes may suffer arbitrary (i.e., \emph{Byzantine}) faults at any time, and $3F+1\leq N$. All remaining nodes are \emph{honest}---they follow the designed routine exactly. 
Specifically, Byzantine nodes may seek to compromise both the consensus (model consistency) and accuracy of the trained model through the following activities: 
\emph{(i) equivocation}---sending conflicting information, such as discrepant model updates or conflicting protocol messages of any stage of the designed routine, to different nodes;
\emph{(ii) model manipulation}---sending arbitrary model updates, including poisonous models crafted through MPAs, to other nodes to influence their local aggregation;
\emph{(iii) non-response}---omitting model updates in the hope of causing honest nodes to wait indefinitely or proceed with incomplete updates.

\emph{Non-goal.} We do not consider privacy attacks in this work, though it is an important orthogonal issue in FL and DFL.
\section{\sysname System Design}
\label{sec:sys-design}


\sysname builds on the core intuition that the decentralized nodes can leverage a consensus protocol to obtain a common subset of at least $N-F$ model updates and use them for local aggregation, achieving model consistency and better Byzantine resilience compared to neighbor-only aggregation. 
Fig.~\ref{fig:sys-arch} illustrates DFL-C workflow in one training round. 

The round begins with \textbf{local training}, where node $i$ retrains model $\theta^t$ using its own data $ds_i$. This yields a local model update $\delta_i$. To establish a consistent and robust input set for aggregation, all nodes participate in the \textbf{T-ACS} consensus by proposing their local updates. T-ACS extends the baseline ACS consensus protocol to allow honest nodes to finalize a common subset of model updates $\{\delta_j\}_{j \in \text{CS}}$ of size $N-F$ or more. T-ACS differs from baseline ACS in that
the order in which the proposed updates are processed is guided by the trust scores from the preceding round. Updates from nodes with higher trust scores are prioritized during message propagation and delivery, expediting the termination of T-ACS.

After T-ACS consensus, each node independently executes a deterministic \textbf{local aggregation} phase that involves computing trust scores for nodes, denoted $\{S_i\}$ for all $i$. We design the \textbf{\tsname} scheme to compute trust scores to penalize Byzantine/unreliable behaviors in both the consensus domain and data domain (model quality). Then, weighted aggregation is used to obtain the new model:
\begin{equation}
    \hat{\theta}^{t+1} = \hat{\theta}^t + \sum_{i \in \text{CS}} S_i \delta_i
\end{equation}

The updated global model $\hat{\theta}^{t+1}$ is stored as a checkpoint and used for the next training round, or inference tasks if the current round completes. Importantly, the trust scores computed by \tsname inform a priority list for the next round's T-ACS consensus, fulfilling the latter's prioritization design.

Next, we describe the detailed design of T-ACS, \tsname, and a temporal heuristic for improving consensus efficiency.

\subsection{Preliminary: Asynchronous Common Subset}
\label{subsec:building-blocks-acs}

ACS \cite{miller2016honey,guo2020dumbo,duan2023fin}
is a type of BFT consensus protocol that enables a group of nodes to agree on a common subset of individually proposed values, even under full asynchrony and up to $F < \frac{1}{3}N$ Byzantine nodes. The key guarantee is that all honest nodes will output at least $N - F \geq \frac{2}{3}N + 1$ consistent proposals.
The ACS protocol typically comprises two stages---Reliable Broadcast (RBC) and {Binary Agreement (BA). 

    \textbf{RBC stage.} Each node $i$ initiates one RBC instance, denoted $RBC_i$, for its own proposal and participates in $N-1$ RBC instances initiated by its peers. RBC proceeds in three phases: VAL---the proposer sends its value to all nodes; ECHO---upon receiving a VAL, each node re-broadcasts it to all others; READY---once sufficient ECHO messages ($N-F$ out of $N$) are received, each node broadcasts a READY and decides delivery.
    RBC guarantees that either all correct nodes eventually deliver the same message, or none do, maintaining consistency under Byzantine behavior.
    
    \textbf{BA stage.} After each RBC delivery, each node encodes whether it received the RBC proposal into a binary vector and enters a BA process. The BA protocol proceeds through multiple rounds of binary voting, where each node exchanges its current binary value and updates it based on quorum thresholds. The result is an agreement on which proposals are valid and should be finalized in the common subset.

\begingroup
\makeatletter
\let\oldalgocfpreruled\@algocf@pre@ruled
\renewcommand{\@algocf@pre@ruled}{%
    \kern3pt%
    \oldalgocfpreruled%
}
\makeatother

\begin{algorithm}
\caption{T-ACS Protocol Pseudocode for Node $j$}
\label{alg:T-ACS}
\KwIn{
Trust scores $\{S_i^t\}_{i \in \mathcal{N}}$, local update $\delta_j^t$
}
\KwOut{Common subset $\{\delta_i^t\}_{i \in CS^t}$}
\tcc{\small Step 1: Define Priority Order}
$\pi^t \gets \text{SortDescending}(\{S_i^t\}_{i \in \mathcal{N}})$ 

\tcc{\small Step 2a: Trust-informed RBC}
count $\gets 0$\;
MsgBuf $\gets []$\;
Broadcast RBC\_VAL($\delta_j^t$)\;
\While{RBC not complete}{
    MsgBuf $\gets$ MsgBuf $\cup$ NewRBCMessages\;
    \textbf{Sort} MsgBuf by $\pi^t$\;
        msg $\gets$ MsgBuf.pop()\;
        ProcessRBC(msg)\;
        \If{msg.type = READY \textbf{and} msg.sender $\in$ top-$\frac{2N}{3}$}{
            count $\gets$ count $+ 1$\;
        }
    \If{count $> \frac{2N}{3}$}{
        start BA; RBC moved to background}
}

\tcc{\small Step 2b: Trust-informed BA}
count $\gets 0$\; 
MsgBuf $\gets []$\;
Broadcast BA\_MSG\;
\While{BA not complete}{
    MsgBuf $\gets$ MsgBuf $\cup$ NewBAMessages\;
    \textbf{Sort} MsgBuf by $\pi^t$\;
        msg $\gets$ MsgBuf.pop()\;
        ProcessBA(msg)\;
        \If{msg.sender $\in$ top-$\frac{2N}{3}$}{
            count $\gets$ count $+ 1$\;
        }
        \If{count $> \frac{2N}{3}$}{
            update $\{\delta_j^t\}_{j \in CS^t}$\;
            continue BA in background;
        }
}
\end{algorithm}
\endgroup

\subsection{T-ACS: Trust-informed ACS Consensus}
\label{subsec:tacs}

To speed up the ACS consensus, we introduce T-ACS, a trust-informed adaptation of the state-of-the-art ACS baseline \cite{duan2023fin}. T-ACS incorporates trust scores to guide the relative prioritization of RBC execution. As shown in Algorithm \ref{alg:T-ACS}, each node maintains the list of proposers from the previous round \( \mathcal{V} \) and their corresponding trust scores \( \{S_j^t\} \). When initiating or processing RBC instances, nodes sort all RBC tasks in descending order of the proposers' trust scores (i.e., per the sorted local priority list \( \pi^t \)). The node then prioritizes message forwarding, echoing, and readiness checks according to this order, ensuring that higher-trust RBCs are more likely to complete early. 
This scheme does not exclude low-trust nodes outright; it adjusts the delivery timing leveraging trust scores as a soft signal.
As a result, this design significantly speed up the common subset consensus finalization by pre-selecting proposals from high-trust nodes for early RBC completion throughout the ACS process. 

\textbf{Equivocation signal for \tsname:} If a node equivocates, i.e., sending RBC messages that cannot be verified as the same original message, this misbehavior is detectable by standard cryptographic signatures in the READY phase. In the baseline ACS scheme \cite{duan2023fin} that T-ACS builds on, RBC messages are erasure-coded for efficiency, so that the non-reconstructability of coded fragments also indicate message equivocation. 

\subsection{\tsname: Dual-Domain Trust Scoring}

\tsname penalizes Byzantine and unreliable behaviors across two operational domains: consensus and data. 
For the consensus domain, we consider \emph{unequivocation} and \emph{non-responsiveness}, which are unique to our consensus-based framework; they are based on whether a node's model update is included in the common subset and equivocation signal provided by the consensus algorithm. For the data domain, we consider \emph{spatial inconsistency} and \emph{temporal instability} as inspired by prior work \cite{wang2022flare,krauss2023mesas}, respectively.

\vspace{2pt}
\noindent
\textbf{\underline{Case 1: Within Common Subset}} If node $i$'s model update is successfully accepted into the common subset after the T-ACS phase, its trust score $S_i$ is calculated as follows.
We first define the spatial consistency score $C_i$ based on Multi-Krum~\cite{blanchard2017machine}, using the average squared Euclidean distance to the $K = N - F - 2$ closest peer models:
\begin{equation}
C_i = \exp \Bigg(-\frac{1}{\gamma_1} \sum_{j \in \mathcal{N}_i^{(K)}} ||\delta_i - \delta_j||_2^2 \Bigg)
\end{equation}
where $\mathcal{N}_i^{(K)}$ is the set of $K$ nearest neighbors of node $i$, and $\gamma_1$ is a scaling constant.
This score promotes alignment with the majority of honest nodes, making it effective for detecting outliers and mitigating the influence of anomalous updates.

Second, we define temporal stability $T_i$ as the inverse of the average update variation over the past $W$ rounds:
\begin{equation}
T_i = \exp\left(-\frac{1}{\gamma_2}\frac{1}{W} \sum_{w=1}^{W} ||\delta_i^{(t-w)} - \delta_i^{(t-w-1)}||_2^2 \right)
\end{equation}
where $\gamma_2$ is another scaling constant. This score discourages model updates that exhibit high variance across rounds. Rapid legitimate distribution shifts may also temporarily reduce $T_i$; therefore, the current temporal metric is most suitable when honest update trajectories evolve sufficiently smoothly over the $W$-round observation window. Adapting D2TS to abrupt distribution shifts is left to future work.

Then, an instant trust score $q_i$ is computed:
\begin{equation}
q_i = (C_i)^{\alpha} \cdot (T_i)^{1-\alpha} 
\end{equation}
where $\alpha$ is an adjustable parameter controlling the relative emphasis on spatial consistency $C_i$ and temporal stability $T_i$. In our evaluation, we set $\alpha=0.5$ as a neutral configuration that gives the two factors equal emphasis without task-specific tuning. A lower $\alpha$ places more emphasis on temporal stability, while a higher $\alpha$ places more emphasis on spatial consistency. We leave systematic sensitivity analysis of $\alpha$ under different data dynamics to future work.

Finally, to ensure trust score continuity across rounds and allow the system to amplify consistent trustworthy behavior gradually, we apply a moving average to derive the new $S_i$:
\begin{equation}
S_i^{(t)} = (1 - \kappa) \cdot S_i^{(t-1)} + \kappa \cdot \frac{q_i}{\sum_{k \in \mathcal{CS}} q_k}
\end{equation}
where $\kappa$ controls the smoothing weight.

\vspace{2pt}
\noindent
\textbf{\underline{Case 2: Equivocation Detected}} When a node is found to have equivocation (as labeled by T-ACS), its model update is not accepted into the common subset, and its trust score will face the equivocation penalty. Specifically:
\begin{equation}
S_i^{(t)} =
\begin{cases}
0 & \parbox[t]{0.7\linewidth}{\raggedright if node $i$ has been accused of equivocating in any of the past $H$ rounds} \\
1 & \text{otherwise}
\end{cases}
\end{equation}
with $H$ being the window size for tracking historical equivocation.
This binary penalty enforces strict accountability by permanently excluding nodes with a history of equivocation from influencing aggregation.

\vspace{2pt}
\noindent
\textbf{\underline{Case 3: Remaining Nodes}} For all remaining nodes that are not included in the common subset nor found for equivocation, the trust score of this node is calculated as follows:
\begin{equation}
\label{eq:ts-remaining}
S_i^{(t)} = (1 - \kappa) \cdot S_i^{(t-1)}
\end{equation}
where $\kappa$ is the same smoothing coefficient used for case 1. 
This design takes inspiration from prior art in asynchronous FL in the classic setting 
\cite{xie2019asynchronous,wang2022asynchronous} that tackles the ``straggler problem'' where the overall training can be bottlenecked by the slowest clients, by weighting fresh updates more heavily.
It applies a gradual decay for late or missing updates rather than an immediate zero score. This gradualism preserves partial trust for nodes affected by temporary network delays, allowing them to restore trust scores by becoming Case-1 nodes in subsequent rounds.

In evaluation, we heuristically set the smoothing coefficient to $\kappa=0.2$, which gives each round's instant score a 20\% weight while retaining 80\% of historical trust. This setting provides conservative trust evolution across rounds; systematic parameter sensitivity and formal trust-score stability under adaptive attacks are left to future work.

\subsection{Adaptive Waiting for Practical Efficiency amid Asynchrony}
\label{subsec:optimization}

While \sysname is designed to tolerate fully asynchronous network conditions, much of the asynchrony in practice is systemic and predictable. Specifically, the heterogeneity of training speeds leads to faster nodes spending excessive time idling in consensus, an overhead that can be practically minimized. We design an \emph{adaptive waiting policy} that leverages the temporal heuristic---the T-ACS protocol's termination events can be used for clocking the subsequent consensus starting points for the whole group.
Fig.~\ref{fig:sys-optimize} illustrates this policy. After local training concludes in round \( t \), each node estimates its preferred delay \( d_i \) before initiating the next consensus based on its compute load and training time, i.e., as its estimated training time for the next round. These delay proposals are embedded as auxiliary data in each node’s RBC message and processed by BA as part of T-ACS.
Once consensus is reached, all honest nodes agree on a common start delay $D^t$ by taking the $r$-th smallest of $\{d_i\}$, where $r$ is a predetermined parameter. 
Each node tracks its consensus period using local time markers. Let \( W_{\text{i,start}}^t \) denote node $i$'s local end time of the current consensus round and also the waiting start point, then node $i$'s scheduled start time of the next round is:
$    W_{\text{i,end}}^t = W_{\text{i,start}}^t + D^t$.
This delay \( D^t \) serves as the system-wide trigger for initiating the next round of T-ACS.

As a result, the adaptive waiting policy reduces the network's cumulative time spent on consensus without incurring additional communication costs or compromising consensus safety. Fast nodes can allocate the freed-up processor time for other operations or enter energy saving.



\begin{figure}
    \centering
    \includegraphics[width=0.48\textwidth]{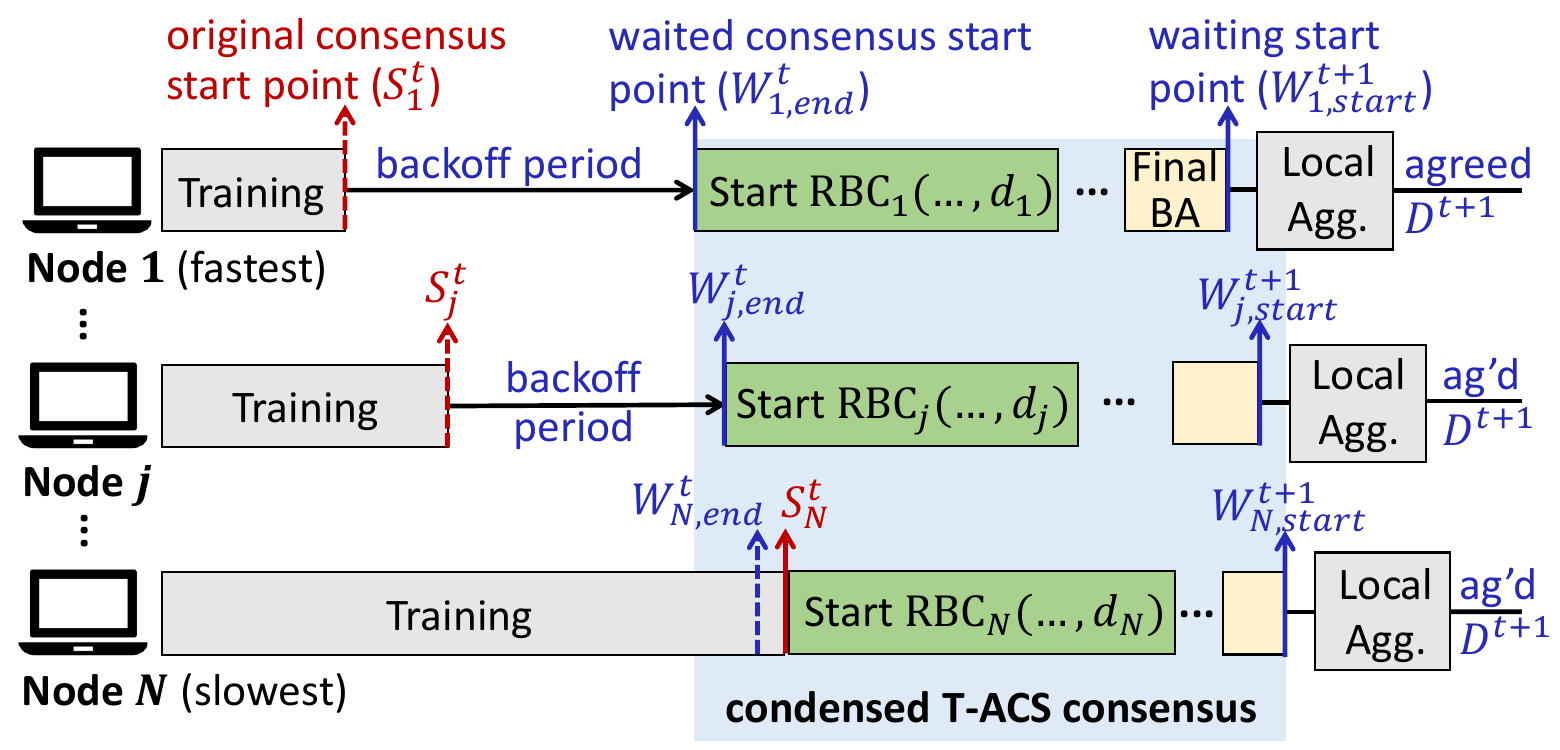}
    \caption{Example of the adaptive waiting policy achieving condensed consensus for \sysname.}
    \label{fig:sys-optimize}
\end{figure}


\begin{figure*}
    \centering
    \begin{subfigure}{0.32\textwidth}
        \centering
        \includegraphics[width=\textwidth]{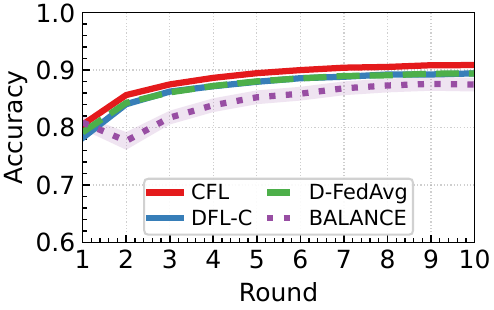}
        \caption{Under Equivocation Attack}
        \label{fig:plt2}
    \end{subfigure}
    \hfill
    \begin{subfigure}{0.32\textwidth}
        \centering
        \includegraphics[width=\textwidth]{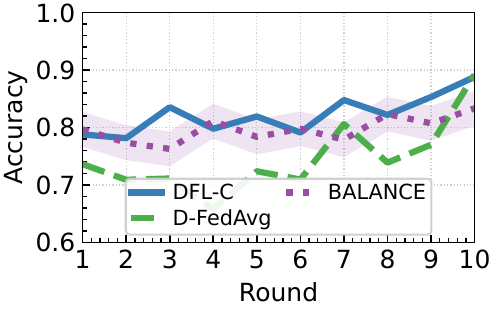}
        \caption{Under Label-Flipping Attack}
        \label{fig:plt3}
    \end{subfigure}    
    \hfill
    \begin{subfigure}{0.31\textwidth}
        \centering
        \includegraphics[width=\textwidth]{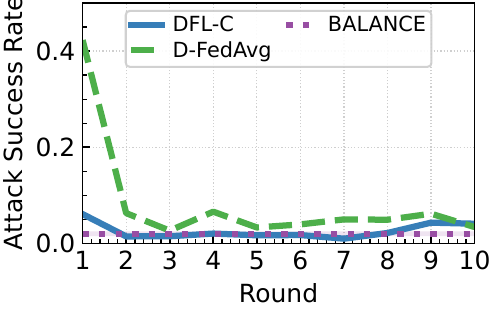}
        \caption{Under Backdoor Attack (Badnets)}
        \label{fig:plt4}
    \end{subfigure}
    \caption{
    Evolution of model training under different Byzantine attacks. The LeNet-5 model and the fMNIST dataset are used. Shaded region indicates standard deviation across nodes in BALANCE \cite{fang2024byzantine} due to model divergence. }
    \label{fig:plt2&3&4}
\end{figure*}

\section{Analyses}
\label{sec:analysis}



Under the threat model in \S\ref{subsec:threat-model}, here we show how \sysname achieves the proposed design objectives.

\begin{proposition}[Global model consistency]
\label{prop:model-consistency}
\sysname guarantees all honest nodes obtain the same global model $\hat{\theta}_{t}$ at the end of a round despite network asynchrony and equivocation. 
\begin{IEEEproof}[Proof Sketch]
This property naturally follows from T-ACS's consensus guarantee---all honest nodes obtain the same common subset $\{\delta_i\}_{i\in CS}$ and use it for local aggregation, leading to the same new global model.
Now consider that $F$ nodes equivocate. According to the design of ACS' RBC phase (\S\ref{subsec:building-blocks-acs}), if a Byzantine node $j$ equivocates its model update (i.e., proposes multiple versions of VAL messages, each containing a different $\theta_j$), honest nodes will send ECHO messages containing different $\theta_j$ to each other. Each honest node does not receive sufficient ECHO messages for $\theta_j$ and thus will not broadcast the READY message for it, making node $j$'s RBC incomplete. As a result, all items in $\{\delta_i\}_{i\in CS}$ come from different nodes and are unequivocal.
\end{IEEEproof}
\end{proposition}


\begin{proposition}[Byzantine manipulation threshold]
\label{prop:mpa-resilience}
\sysname ensures that each node aggregates over model updates that have an honest majority, i.e., fewer than 50\% are malicious.
\begin{IEEEproof}[Proof Sketch]
After each ACS consensus in a federated training round, the resulting common subset $\{\delta_i\}_{i\in CS}$ includes at least $N - F$ model updates from all nodes. Assuming $N \geq 3F + 1$, the proportion of Byzantine updates in this subset is bounded by $\frac{F}{N - F} < \frac{1}{2}$, ensuring an honest majority during the subsequent aggregation step.
\end{IEEEproof}
\end{proposition}

\textbf{Remark:} This honest majority threshold is commonly used by existing Byzantine-robust aggregation methods for classic FL \cite{blanchard2017machine,yin2018byzantine} as a precondition.


\textbf{Complexity.}
Compared to existing DFL, the main additional overhead of DFL-C comes from the T-ACS consensus executed in every training round. In theory, one T-ACS run incur $O(N^2)$ messages for each node due to all-to-all interactions among $N$ proposers. 
However, recent optimized ACS primitives such as FIN~\cite{duan2023fin}, which serves as our baseline, show that ACS can achieve $O(1)$ expected clock time while still despite the $O(N^2)$ message complexity, because they avoid the classical design that requires $N$ parallel BA instances and instead use a succinct procedure that uses only an expected constant number of
BA agreement instances. T-ACS inherits this optimization and make further improvement: it reduces practical consensus latency by prioritizing higher-trust proposals and shortening the critical path to termination. 
Despite this reduction, consensus latency still increases with model payload size, as shown in §VI-B. Moreover, the all-to-all communication pattern makes communication cost grow rapidly with $N$. Our current prototype evaluation therefore establishes practicality primarily for relatively small collaborative networks up to the evaluated scale of $N=13$; scalability to substantially larger networks remains future work.

\section{Implementation and Evaluation}
\label{sec:implementation}

We implemented a proof-of-concept DFL-C system in roughly 1200 lines of Python code.\footnote{Available at \url{https://github.com/yli568/DFL-C}.}
For benchmark comparison, we also implemented D-FedAvg, a decentralized version of FedAvg in the DFL-C's consensus framework but without \tsname, and BALANCE \cite{fang2024byzantine}, the state-of-the-art Byzantine-resilient peer-to-peer DFL scheme that does not maintain global model consistency.

\textbf{Experimental Setup.}
We conducted comprehensive experiments to evaluate \sysname's performance in two aspects: (i) accuracy of the trained model under varying network Byzantine influence settings, and (ii) consensus latency overhead attributed to the T-ACS consensus process.
We ran experiments on two platforms: a local computer cluster with GPU nodes, and a group of Raspberry Pi (RPi) 4. 
We deployed one independent \sysname instance per node on both platforms.
Each node runs an independent \sysname program executing the full training pipeline through a round. 
We evaluated two families of neural network-based classifiers: LeNet~\cite{lecun1998gradient} and ResNet~\cite{he2016deep}, on Fashion-MNIST (fMNIST)~\cite{xiao2017fashion} and CIFAR-10~\cite{krizhevsky2009learning} datasets. 
Specifically, we ran LeNet-5 and ResNet-50 on GPU nodes and ran ResNet-18 on the RPi.

\textbf{Attack Settings.} The experimental topology is configured with 4 to 13 nodes (=$N$), and up to $\lceil\frac{N}{3}\rceil$ of them are designated to simulate Byzantine behaviors.
Specifically, Byzantine nodes can perform equivocation attacks (proposing arbitrarily different updates to others) and two widely known MPAs: label-flipping \cite{tolpegin2020data}, an untargeted attack that flips the training labels of selected Byzantine nodes to incorrect classes, 
and Badnets~\cite{gu2019badnets}, a backdoor attack that injects trigger patterns into a small set of training samples and assigns them a fixed incorrect label, causing the global model to misclassify any input containing the trigger towards a target class.

\subsection{Data-plane Performance and Comparison} 
\label{sec-Evaluation}

We first evaluate \sysname's resilience against different types of Byzantine behaviors
in comparison to CFL, D-FedAvg, and BALANCE. The fMNIST dataset is used and $N$ is fixed to 10. D-FedAvg uses the same T-ACS consensus for deriving common subsets, while BALANCE operates under a peer-to-peer topology with each node receiving model updates from 5 neighbors. We enforce a localized Byzantine ratio---at most 1/3 neighbors (i.e., 1 out of 5)---for BALANCE, in line with its original design. 
We also run a centralized FL scheme (CFL) as an ideal benchmark when there is no MPA.

Fig.~\ref{fig:plt2&3&4} shows model performance evolution over 10 training rounds under each attack. Under the equivocation attacks (Fig.~\ref{fig:plt2}),
DFL-C and D-FedAvg converge to accuracy levels close to CFL, while DFL-C additionally preserves a common model across honest nodes.
Under label-flipping attacks (Fig.~\ref{fig:plt3}),
DFL-C maintains competitive accuracy relative to BALANCE and clearly outperforms D-FedAvg, while preserving global consistency. 
Under the backdoor attack, Byzantine nodes start attacking in the first round.
Fig.~\ref{fig:plt4} shows the attack success rate (ASR) of the backdoor attack. Both DFL-C and BALANCE effectively suppress the ASR to near zero across all rounds, whereas D-FedAvg suffers from significantly higher ASR on the first round, indicating vulnerability to backdoor injection.

\begin{figure*}
    \centering
    \begin{subfigure}{0.32\textwidth}
        \centering
        \includegraphics[width=\textwidth]{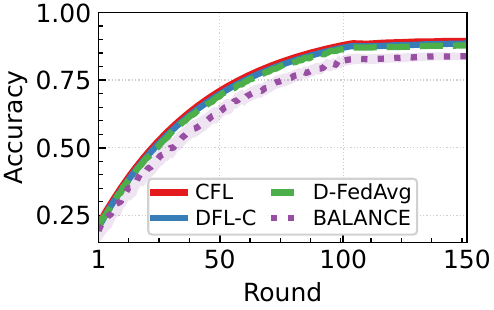}
        \caption{Under Equivocation Attack}
        \label{fig:plt10}
    \end{subfigure}
    \hfill
    \begin{subfigure}{0.32\textwidth}
        \centering
        \includegraphics[width=\textwidth]{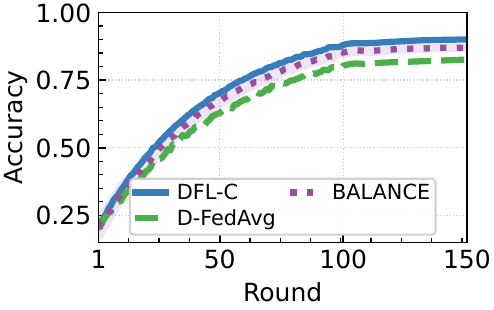}
        \caption{Under Label-Flipping Attack}
        \label{fig:plt11}
    \end{subfigure}    
    \hfill
    \begin{subfigure}{0.31\textwidth}
        \centering
        \includegraphics[width=\textwidth]{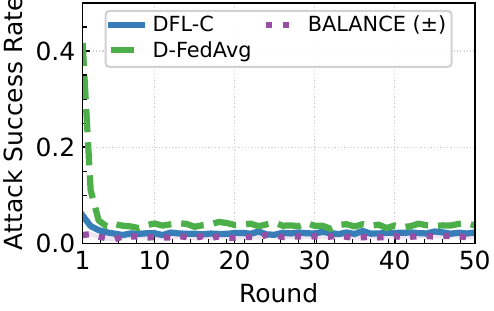}
        \caption{Under Backdoor Attack (Badnets)}
        \label{fig:plt12}
    \end{subfigure}
    \caption{Evolution of model training under different Byzantine attacks. The ResNet-50 model and the CIFAR-10 dataset are used based on the incremental data feed setting.
    }
    \label{fig:plt10&11&12}
\end{figure*}

To validate that these trends hold under a more demanding workload, we repeat the same attacks on CIFAR-10 using ResNet-50 on local GPU nodes. We adopt an incremental data feed setting to emulate real-world data arrivals. At each round, every node receives an additional $1/100$ fraction of its local training data; it observes the full local dataset after $100$ rounds. 
The result is shown in Fig.~\ref{fig:plt10&11&12}. Over longer training horizons, DFL-C continues to track the centralized baseline under equivocation and maintains a clear accuracy advantage over D-FedAvg under label-flipping. Under Badnets, DFL-C keeps the attack success rate near zero throughout training, while D-FedAvg exhibits a noticeably higher ASR early on.

Since DFL-C couples robust aggregation with common-subset consensus, all honest nodes aggregate over the same agreed update set and can explicitly filter equivocation. This design targets model consistency and equivocation resilience while preserving competitive task accuracy.

\textbf{Performance Under Non-IID Scenarios.} Table~\ref{tab:dirichlet} summarizes the trained LeNet-5 accuracy under IID and non-IID data scenarios for both fMNIST and CIFAR-10, across different system scales ($N\in\{7,10\}$), with and without Byzantine nodes (label-flipping). We used the Dirichlet distributions to emulate non-IID cases: we generate label-skewed partitions, for each class, a client-allocation probability vector from a symmetric Dirichlet distribution.
The concentration parameter $\alpha$ controls statistical heterogeneity (i.e., non-IID degree): larger $\alpha$ makes the per-class allocation more uniform across nodes and therefore closer to IID; smaller $\alpha$ concentrates each class on fewer nodes and thus more non-IID. 

\begin{table}

    \vspace{5pt}
    \caption{Comparing Accuracy of DFL-C and BALANCE \cite{fang2024byzantine} at convergence when training LeNet-5 under IID and Dirichlet distribution-based non-IID (nIID) settings.}
    \label{tab:dirichlet}
    \footnotesize
    \setlength{\tabcolsep}{3.5pt}
    \begin{tabular}{l|c|c|c}
    \toprule
         & N=7 Non-Byz & N=10 Non-Byz & N=10 Byz \\
    \midrule
    \rowcolor[gray]{0.9}
    \multicolumn{4}{l}{\textbf{DFL-C}} \\
        fMNIST (IID)          & 0.8955 & 0.8905 & 0.8863 \\
        -- (nIID., $\alpha$=$1.0$) & 0.7743 & 0.7985 & 0.7922 \\
        -- (nIID, $\alpha$=$0.3$) & 0.6380 & 0.6941 & 0.6829 \\
        CIFAR-10 (IID)        & 0.5677 & 0.5689 & 0.5593 \\
        -- (nIID, $\alpha$=$1.0$) & 0.5072 & 0.5284 & 0.5117 \\
        -- (nIID, $\alpha$=$0.3$) & 0.4398 & 0.4812 & 0.4568 \\
    \midrule
    \rowcolor[gray]{0.9}
    \multicolumn{4}{l}{\textbf{BALANCE}} \\
        fMNIST (IID)                & 0.86 [0.82, 0.89] & 0.86 [0.81, 0.89] & 0.82 [0.75, 0.87] \\
        -- (nIID, $\alpha$=$1.0$) & 0.74 [0.64, 0.81] & 0.77 [0.66, 0.84] & 0.71 [0.57, 0.81] \\
        -- (nIID, $\alpha$=$0.3$) & 0.58 [0.43, 0.68] & 0.63 [0.47, 0.74] & 0.55 [0.35, 0.69] \\
        CIFAR-10 (IID)              & 0.54 [0.49, 0.57] & 0.55 [0.49, 0.58] & 0.51 [0.43, 0.56] \\
        -- (nIID, $\alpha$=$1.0$) & 0.49 [0.39, 0.54] & 0.51 [0.41, 0.57] & 0.45 [0.32, 0.54] \\
        -- (nIID, $\alpha$=$0.3$) & 0.40 [0.27, 0.47] & 0.43 [0.29, 0.51] & 0.36 [0.19, 0.48] \\
    \bottomrule
    \end{tabular}

    \vspace{4pt}
    Note: For BALANCE, due to local model discrepancy across the nodes, each data entry shows a three-tuple: average accuracy [worst-node accuracy, best-node accuracy].
\end{table}

The results shows three trends for \sysname. First, IID consistently achieves the highest accuracy, while stronger non-IID reduces accuracy on both datasets, confirming the cost of statistical skew.
Second, under the non-IID settings, increasing the number of nodes from 7 to 10 generally improves accuracy, suggesting that wider participation partially mitigates local data skew through broader aggregate coverage. 
Third, introducing Byzantine nodes causes only a modest additional drop relative to the corresponding non-Byzantine setting. Also, CIFAR-10 remains harder for LeNet-5 than fMNIST, so its absolute accuracy is lower across all partition settings.

Notably, \sysname achieves significantly better accuracy than BALANCE when the non-IID degree increases and there are Byzantine nodes. In BALANCE, since updates each node only from its own and accepted neighbor models, 
different nodes may observe substantially different effective training distributions and therefore exhibit a noticeable accuracy spread. This effect becomes especially pronounced when a node's neighbors all hold strongly non-IID data, and in the worst case, a large fraction of neighbors are Byzantine nodes.
By contrast, DFL-C mitigates this locality-induced degradation by enforcing a common subset of updates for uniform aggregation across honest nodes.

\subsection{Consensus Latency Overhead}
\label{subsec:eval-acs}

\begin{figure}[ht]
    \centering
    \includegraphics[width=0.9\linewidth]{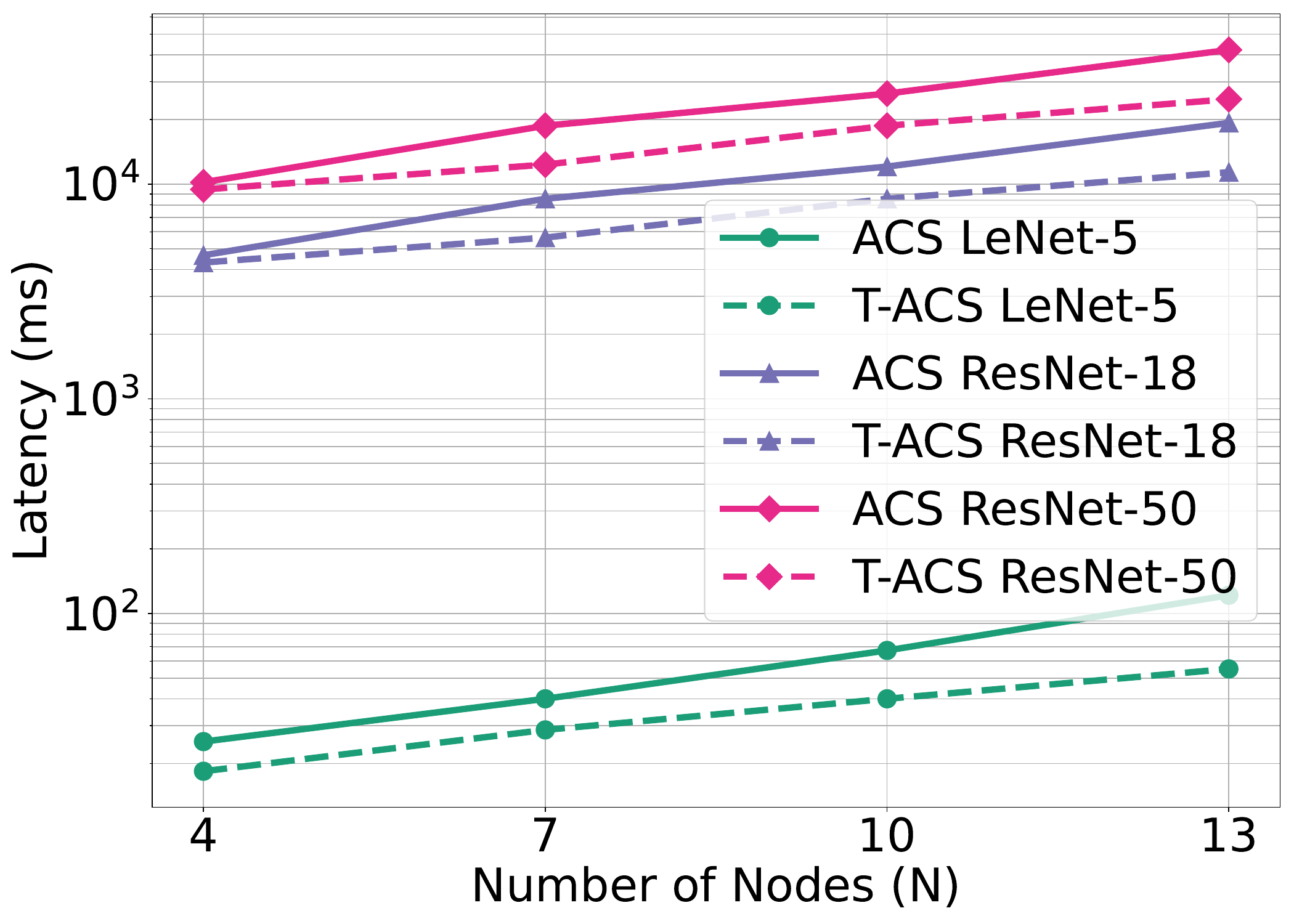}
    \caption{Consensus latency of T-ACS and the baseline ACS protocol \cite{duan2023fin} carrying different model payloads.}
    \label{fig:plt13}
\end{figure}

The primary measured runtime overhead introduced by DFL-C compared to existing DFL schemes lies in its T-ACS consensus mechanism.
Fig.~\ref{fig:plt13} shows the consensus latency of baseline ACS \cite{duan2023fin} and \sysname's T-ACS when carrying model payload sizes comparable to LeNet-5, ResNet-18, and ResNet-50, under varying node counts ($N=4$, $7$, $10$, and $13$). 
As expected, consensus latency increases with both the model payload size and $N$. 
Across all model families and system scales, T-ACS consistently achieves lower latency than ACS, and the reduction becomes more pronounced as $N$ grows. 
In particular, T-ACS stays at the tens-of-milliseconds level for LeNet-5-scale payloads, while incurring seconds to tens-of-seconds latency for ResNet-scale payloads. Despite the growing consensus time with $N$ and payload size, it is worth noting that
in all of our experiments, the consensus runtime is minimal compared to local training time. On GPU nodes, one training session takes 3--10\,s. while on Raspberry Pi CPUs, one round takes 1--4\,hours.

\subsection{Effect of Adaptive Waiting Policy}
\label{subsec:eval-waiting-policy}

To quantify the efficiency benefits of the adaptive waiting policy described in \S\ref{subsec:optimization}, we simulate scenarios where node training times differ so that the generation of model updates exhibits a degree of asynchrony. We configure each node independently to sample its training time from a uniform distribution $\mathcal{U}(T, T \cdot (1 + \Delta))$, where $T$ is the base training time and $\Delta$ represents the heterogeneity levels ranging from 10\% to 50\%. The synchronized delay $D^t$ is set to the training time of the $\lceil \frac{2N}{3} \rceil$-th fastest node.
Table~\ref{tab:adaptive-saving} shows the average idle time saved by the adaptive waiting policy under $\Delta$.

\begin{table}[ht]
    \caption{Average per-node time saved}
    \centering
    \begin{tabular}{cccccc}
    \toprule
    $\Delta$ & 10\% & 20\% & 30\% & 40\% & 50\% \\
    \midrule
    $N=7$  & 3\%  & 6\%  & 10\% & 13\% & 16\% \\
    $N=11$ & 4\%  & 8\%  & 11\% & 15\% & 20\% \\
    \bottomrule
    \end{tabular}
    \label{tab:adaptive-saving}
\end{table}

The result shows that this policy saves idle time for nodes, with per-node savings increasing as heterogeneity grows, showing the feasibility of the adaptive waiting scheme.

\section{Conclusion}

We present \sysname, a novel Byzantine-resilient decentralized federated learning architecture to enable the collaborative training of a common model in mission-critical scenarios. By integrating common subset consensus and dual-domain trust scoring, \sysname obtains model consistency and mitigates equivocation in an asynchronous network with Byzantine participants.
\sysname harmonizes trust scores to inform a more efficient consensus and leverage a temporal heuristic to further improve time efficiency amid systemic computational heterogeneity.
Our evaluation shows that DFL-C maintains robust model accuracy in the presence of Byzantine nodes across diverse settings while guaranteeing model consistency: it achieves higher accuracy than the state of the art under untargeted model poisoning and comparable resilience against backdoor attacks, with the advantage widening as the data becomes more non-IID. The results also demonstrate the efficacy of trust score-informed consensus and of the practical temporal heuristic in improving time efficiency. Future work includes more efficient consensus mechanism design for larger networks and heavier model payloads, and systematic D2TS parameter sensitivity analysis under adaptive attacks.

\section*{Acknowledgment}
This work was supported in part by the Office of Naval Research under grant N00014-24-1-2730 and the US National Science Foundation under grant 2442382.

\bibliographystyle{ieeetr}
\bibliography{bib-FL,bib-DFL,bib-AsyncFL,bib1,reference}

\end{document}